\documentclass[10pt,conference]{IEEEtran}
\IEEEoverridecommandlockouts
\usepackage{orcidlink} %
\usepackage{graphicx} %
\usepackage{float}
\usepackage{amsmath} %
\usepackage{amssymb}  %
\usepackage{mathtools}
\usepackage{bbm}
\usepackage{nicefrac}

\usepackage[utf8x]{inputenc}
\usepackage[T1]{fontenc}
\usepackage[10pt]{moresize}
\usepackage{eurosym}
\usepackage{listings}
\usepackage{fontawesome}
\usepackage{color}
\usepackage{booktabs}
\usepackage{adjustbox}
\usepackage{colortbl}
\usepackage{tabularx}
\usepackage{url}  %

\usepackage{tikz}
\usepackage{subcaption}

\usepackage{multirow}
\usepackage{csvsimple}
\usepackage{array}
\usepackage[usestackEOL]{stackengine}
\usepackage{siunitx}
\usepackage{etoolbox}
\usepackage[numbers]{natbib} 
\usepackage[frozencache=true,cachedir=mintedcache]{minted}
\makeatletter
\newcommand{\currentfontsize}{\fontsize{\f@size}{\f@baselineskip}\selectfont}
\makeatother
\setmintedinline{fontsize=\currentfontsize}

\usepackage{paralist}

\renewcommand{\descriptionlabel}[1]{\textbf{\textsc{#1}}}

\usepackage{import}
\usepackage{algorithm}
\usepackage{algpseudocode}

\usepackage{hyperref}
\usepackage[capitalise,nameinlink]{cleveref}
\usepackage{acro}
\acsetup{
    barriers/use=true
}

\usepackage{todonotes}
\usepackage{wasysym} %

\usepackage{amsthm}
\usepackage{tcolorbox}
\tcbuselibrary{theorems}
\usepackage{framed}
\usepackage{soul} %

\definecolor{t_gray}{HTML}{888888}
\definecolor{t_blue}{HTML}{355fb3}
\definecolor{t_red}{HTML}{b33535}
\definecolor{t_green}{HTML}{3bb335}
\definecolor{t_yellow}{HTML}{b39735}
\definecolor{t_darkgray}{HTML}{454545}
\definecolor{t_darkblue}{HTML}{1e3666}
\definecolor{t_darkgreen}{HTML}{22661e}
\definecolor{t_darkred}{HTML}{661e1e}
\definecolor{t_darkyellow}{HTML}{66571e}
\definecolor{t_lightblue}{HTML}{8ea7d7}
\definecolor{t_lightred}{HTML}{dc8989}
\definecolor{t_lightgreen}{HTML}{8ddc89}

\definecolor{verylightgray}{RGB}{240,240,240}

\newcommand{\geiger}{\textsc{go-geiger}}

\newcommand{\ourtool}{\textsc{UnGoML}}
\newcommand{\product}{\ourtool}
\newcommand{\codebert}{\textsc{CodeBERT}}
\newcommand{\lOne}{\texttt{WHAT}}
\newcommand{\lTwo}{\texttt{WHY}}

\newcommand{\footurl}[1]{\footnote{\url{#1}}}

\newcommand{\code}[1]{\mintinline{go}{#1}}
\newcommand{\unsafe}{\texttt{unsafe}}

\newcommand*{\condbold}[3][]{\ifthenelse{\equal{#2}{1}#1}{{\mathbf{#3}}}{#3}}
\newcommand*{\condline}[3][]{\ifthenelse{\equal{#2}{1}#1}{{\underline{#3}}}{#3}}
\newcommand*{\condgreen}[3][]{\ifthenelse{\equal{#2}{1}#1}{{\textcolor{t_darkgreen}{#3}}}{#3}}
\newcommand*{\condblue}[3][]{\ifthenelse{\equal{#2}{1}#1}{{\textcolor{t_blue}{#3}}}{#3}}
\newcommand*{\condred}[3][]{\ifthenelse{\equal{#2}{1}#1}{{\textcolor{t_red}{#3}}}{#3}}
\newcommand*{\evalres}[6][]{%
    \ifthenelse{\equal{#3}{m}}{OOM}{%
    \ifthenelse{\equal{#3}{t}}{OOT}{%
        $\condred[\and\equal{#3}{0}]{#4}{\condblue[\and\equal{#4}{0}]{#3}{\condbold[#1]{#2}{\num{#5}}}} \pm \num{#6}$%
    }}%
}

\newcommand*{\evalperc}[6][]{%
        $\num{#5}$ %
}

\newcommand*{\evalsize}[6][]{%
        $\num{#5} \pm \num{#6}$%
}

\newcommand{\covered}{\textcolor{t_darkgreen}{\CIRCLE}}
\newcommand{\notcovered}{\textcolor{t_darkred}{\Circle}}
\newcommand{\partlycovered}{\textcolor{t_darkgreen}{\LEFTcircle}}

\newtcbtheorem{rqbox}{Answer to Research Question}%
{colback=t_blue!5,colframe=t_blue,fonttitle=\bfseries}{th}
\newtcolorbox{summarybox}[1]{colback=t_blue!5!white, colframe=t_blue,fonttitle=\bfseries, title={#1}}

\newcommand{\remMSR}[1]{}

\makeatletter
\newcommand{\linebreakand}{%
  \end{@IEEEauthorhalign}
  \hfill\mbox{}\par
  \mbox{}\hfill\begin{@IEEEauthorhalign}
}
\makeatother

\hypersetup{
  colorlinks = false,
  linkbordercolor  = t_blue,
  citebordercolor  = t_darkgreen, 
  urlbordercolor   = t_blue,
}

\newcommand{\dac}[3]{\DeclareAcronym{#1}{short = #2, long = #3}}

\dac{nn}{NN}{neural network}
\dac{lm}{LRM}{logistic regression model}
\dac{svm}{SVM}{support vector machine}
\dac{mlp}{MLP}{multilayer perceptron}
\dac{cnn}{CNN}{convolutional neural network}
\dac{ltr}{LtR}{Learning to Rank}

\dac{gcr}{GC/GR}{graph classification and regression}
\dac{gc}{GC}{graph classification}
\dac{gr}{GR}{graph regression}
\dac{gk}{GK}{graph kernel}
\dac{gnn}{GNN}{graph neural network}
\dac{gcnn}{GCNN}{graph convolutional neural network}
\dac{gcn}{GCN}{graph convolutional network}
\dac{sage}{GraphSAGE}{\textsc{sa}mple and aggre\textsc{g}at\textsc{e}}
\dac{gin}{GIN}{graph isomorphism network}
\dac{wl2gnn}{2-WL-GNN}{2-dimensional Weisfeiler-Lehman GNN}
\dac{sagpool}{SAGPooling}{self-attention graph pooling}
\dac{sampool}{SAMPooling}{softmax attention mean pooling}
\dac{wlst}{WL\textsubscript{ST}}{Weisfeiler-Lehman subtree kernel}
\dac{wlsp}{WL\textsubscript{SP}}{Weisfeiler-Lehman shortest-path kernel}

\dac{ml}{ML}{machine learning}
\dac{xai}{XAI}{explainable artificial intelligence}
\dac{nlp}{NLP}{natural language processing}
\dac{gi}{GI}{graph isomorphism}
\dac{wl}{WL}{Weisfeiler-Lehman}
\dac{ft}{FT}{Fourier transform}
\dac{bfs}{BFS}{breadth-first-search}
\dac{lcm}{LCM}{lowest common multiple}
\dac{dp}{DP}{discriminative power}
\dac{cp}{CP}{computational power}
\dac{ndcg}{NDCG}{normalized discounted cummulative gain}
\dac{mse}{MSE}{mean squared error}
\dac{giq}{GIQ}{generalized inverse quantile}

\dac{gpu}{GPU}{graphics processing unit}
\dac{cfg}{CFG}{control-flow graph}
\dac{ast}{AST}{abstract syntax tree}

\def\BibTeX{{\rm B\kern-.05em{\sc i\kern-.025em b}\kern-.08em
    T\kern-.1667em\lower.7ex\hbox{E}\kern-.125emX}}
\begin{document}

\title{\product{}: Automated Classification of \newline\unsafe{} Usages in Go}

\author{\IEEEauthorblockN{Anna-Katharina Wickert \,\orcidlink{0000-0002-1441-2423}}
\IEEEauthorblockA{\textit{Software Technology Group} \\
\textit{Technische Universität Darmstadt} \\
Darmstadt, Germany\\
wickert@cs.tu-darmstadt.de}
\and
\IEEEauthorblockN{Clemens Damke \,\orcidlink{0000-0002-0455-0048}}
\IEEEauthorblockA{\textit{Institute of Informatics} \\
\textit{University of Munich}\\
Munich, Germany\\
clemens.damke@ifi.lmu.de}
\and
\IEEEauthorblockN{Lars Baumgärtner \,\orcidlink{0000-0002-5805-2773}}
\IEEEauthorblockA{\textit{Software Technology Group} \\
\textit{Technische Universität Darmstadt} \\
Darmstadt, Germany\\
baumgaertner@cs.tu-darmstadt.de}
\linebreakand
\IEEEauthorblockN{Eyke Hüllermeier \,\orcidlink{0000-0002-9944-4108}}
\IEEEauthorblockA{\textit{Munich Center for Machine Learning} \\
\textit{Institute of Informatics, University of Munich}\\
Munich, Germany\\
eyke@lmu.de}
\and
\IEEEauthorblockN{Mira Mezini \,\orcidlink{0000-0001-6563-7537}}
\IEEEauthorblockA{\textit{Hessian Center for Artificial Intelligence (hessian.AI)} \\
\textit{National Research Center for Applied Cybersecurity ATHENE} \\
\textit{Software Technology Group, Technische Universität Darmstadt} \\
Darmstadt, Germany\\
mezini@cs.tu-darmstadt.de}

}

\maketitle

\begin{abstract}
The Go programming language offers strong protection from memory corruption. 
As an escape hatch of these protections, it provides the \unsafe~package.
Previous studies identified that this \unsafe~package is frequently used in real-world code for several purposes, e.g., serialization or casting types.
Due to the variety of these reasons, it may be possible to refactor specific usages to avoid potential vulnerabilities. 
However, the classification of \unsafe~usages is challenging and requires the context of the call and the program's structure. 
In this paper, we present the first automated classifier for \unsafe~usages in Go, \ourtool, to identify \emph{what} is done with the \unsafe~package and \emph{why} it is used. 
For \ourtool{}, we built four custom deep learning classifiers trained on a manually labeled data set. 
We represent Go code as enriched \acp{cfg} and solve the label prediction task with one single-vertex and three context-aware classifiers. 
All three context-aware classifiers achieve a top-1 accuracy of more than 86\% for both dimensions, \lOne{} and \lTwo{}.
Furthermore, in a set-valued conformal prediction setting, we achieve accuracies of more than 93\% with mean label set sizes of 2 for both dimensions.
Thus, \ourtool{} can be used to efficiently filter \unsafe{} usages for use cases such as refactoring or a security audit. \\
\textbf{\ourtool:} \url{https://github.com/stg-tud/ungoml} \\
\textbf{Artifact:} \url{https://dx.doi.org/10.6084/m9.figshare.22293052}

\end{abstract}


\section{Introduction}
\label{sec:intro}

In November 2022, NSA released guidance on how to avoid memory vulnerabilities, such as buffer overflows, as these still occur very frequently in code~\cite{nsa2022memorysaftey}.
One of the recommendations is to use
modern programming languages, such as Java, Rust, and Go, with automatic memory management to avoid these vulnerabilities. 
But they also provide escape hatches, such as the \unsafe{} API,
for several purposes.
For instance, empirical studies on the usage of Go's \unsafe{} API in-the-wild~\cite{lauinger2020uncovering,costa2021breaking} show that \unsafe{} is used for calling C libraries, making system calls, serialization, performance optimizations, or for modeling missing language features such as generics\footnote{The latter have been introduced recently, in Go version 1.18, March 2022.}. 

Such escape hatches may, however, reintroduce security risks similar to those in memory unsafe languages~\cite{lauinger2020uncovering, mastrangelo2015use,astrauskas2020how, jung_rustbelt_2018}. 
Hence, their usage requires special attention by software quality teams and developers and should be used rarely. 
However, in practice the usage is wide-spread across Go projects~\cite{lauinger2020uncovering} and occur more frequently than expected by Go experts~\cite{costa2021breaking}. 
Lauinger et al.~\cite{lauinger2020uncovering} analyzed 343 top-rated GitHub projects for two potentially exploitable usage patterns and identified 60 usages of this pattern.
The following report revealed a fix and response rate of over 70~\%~\cite{lauinger2020uncovering}. 
Further, the Go community actively engages and educates about the risks, e.g., in blog posts or talks at their main conference\footnote{\label{gh_ungoml}We added details to the README-file of \ourtool: \url{https://github.com/Cortys/unsafe_go_study_results}.}.

To mitigate the risk, auditing the usages and refactoring to safer alternatives should be considered.
The \unsafe{} API as well as the language evolves, and one may want to perform a large-scale refactoring to introduce the safer alternatives into the code base. 
For example, Generics were recently introduced in Go and can replace the \unsafe{} usages that served to model generics. 
Also, one might want to check if \unsafe{} usages for performing serialization can be replaced with one of the many libraries for efficient serialization. 
Unfortunately, some \unsafe{} usages are unsuitable for large-scale refactoring.
Thus,
one may want to start an in-depth security audit.
A simple search to identify unsafe usages that are flagged as false positives for a static analyzer, reveals that over 370 \unsafe{} usages on GitHub are explicitly marked as audited\footnote{See footnote 2.}.

A prerequisite for such "hardening" actions is efficiently getting an overview of \emph{what} \unsafe{} is used for, e.g., to perform pointer arithmetic or to access memory to compare two addresses, or \emph{why}, e.g., for improving efficiency or for modeling generics. 
Reasoning manually about the \emph{what} and \emph{why} of \unsafe{} usages in large-scale software repositories is cumbersome and time-consuming. 
At the same time, automated inference of the \emph{what} and \emph{why} of \unsafe{} usages
is often challenging to be precisely modeled as rules for all \unsafe{} usage patterns for traditional static analyzers. %
Currently, the static analyzers
only support a small subset of well-studied security-critical \unsafe{} usages~\cite{lauinger2020uncovering, go_vet, go_cmdcompile}. 
Machine learning methods, in general, and deep learning models, in particular, are typically employed to enable the automated handling of complex problems, for which precise modeling is not feasible. 
Such methods are used for various software engineering tasks, e.g., for vulnerability detection~\cite{wang_combining_2021, sonnekalb_deep_2022, fu2022linevul}. 

The work presented in this paper proposes that modern machine learning classifiers, e.g., \acp{gnn}, are well-suited for the automatic classification of the \emph{what} and \emph{why} of \unsafe{} usages -- by considering the context of the calls and the structure of the broader program they have the potential to derive meaningful classifications similar to humans.
We validate this proposition by designing and implementing \ourtool{} -- a tool support quality assurance teams and auditors to obtain an overview of the tasks solved with \unsafe{}. 
\ourtool{} builds upon a large set of manually labeled data of \unsafe{} usages in real-world code~\cite{lauinger2020uncovering} and classifies any \unsafe{} usage along two dimensions, namely \emph{what is done} and \emph{why} \unsafe{} is used. 

We investigated three different \ac{gnn} architectures, more specifically, DeepSets~\cite{Zaheer2017}, \acp*{gin}~\cite{Xu2018}, and the higher-order \acs{wl2gnn} architecture~\cite{Damke2020}. 
A \ac{gnn} architecture seems natural given that code is typically represented in graphs, and recent work on vulnerability classification has shown that \acp{gnn} can improve the accuracy~\cite{chakraborty2021deep, cheng2021deepwukong, zhou2019devign} over token-based approaches. 
The different \ac{gnn} models use the control-flow and edge information differently, e.g., the DeepSets classifier~\cite{Zaheer2017} ignores the control-flow structure and variable usages between the input vertices. 
This enables us to assess the importance of structural information. %

Finally, to effectively support developers and auditors even for ambiguous predictions, we investigate conformal prediction~\cite{Romano2020}, a framework for reliable prediction that comes with statistical guarantees. 
This technique allows for predicting a set of candidate labels covering the true classification for each usage with a prespecified probability, e.g., 95\%. 
Thus, we can present a set of varying sizes instead of a fixed one.
Such sets reflect the fact that sometimes even human annotators have difficulties agreeing on one specific label~\cite{costa2021breaking}.
Further, we assume that conformal set prediction can improve the usability for an end user over the top-1 or top-3 accuracy. The underlying assumption is that the conformal set prediction finds the balance between achieving high accuracy while keeping the set of potential labels as small as possible.
To the best of our knowledge, this work is the first to explore conformal set prediction~\cite{Romano2020} for
software engineering tasks. 

Our results positively validate our proposition that machine learning models are well-suited to classify automatically the what and why of \unsafe{} usages. 
For all three context-aware models, we achieve a combined (\lOne{} and \lTwo{}) top-1 accuracy of nearly 80\% and a top-3 accuracy of over 91\%. 
For the single-vertex model \ac{mlp}, the top-1 accuracy drops to 74\%. 
This indicates that the context is relevant for classifying \unsafe{} usages. 
Further, the average sizes of the sets predicted by the conformal set prediction are around 2 elements with a combined accuracy of over 93\%.  
Thus, conformal set predictions are suitable for \unsafe{} classification to improve the accuracy at the cost of a flexible but on average small set of labels. 

We integrated our prediction models along with \geiger~\cite{lauinger2020uncovering}, into a tool to identify \unsafe{} usages along with the classification of 
any usages of \unsafe{} in a given Go project as the basis for further refactoring or security auditing.

\noindent
In summary, we make the following contributions:
\begin{itemize}
    \item A formalization of the problem of inferring programmers intentions when using \unsafe{} (\lOne{} and \lTwo{}) as a classification problem.
    \item A comparison of relational and non-relational machine learning models, such as \acp{gnn} and DeepSets, to understand the impact of the call graphs on the prediction. 
    \item A discussion of the important features needed to classify the what and why of \unsafe{} usages. 
    \item \ourtool{}, the first classification tool that predicts each \unsafe{} usage within Go projects to understand how and why \unsafe{} is used to support developers and auditors in effectively filtering and judging \unsafe{} usages. 
    \item Initial evidence that conformal set prediction seems valuable and worth exploring for software engineering tasks that leverage classifiers.
\end{itemize}

\begin{table*}[t]
    \caption{An overview of different \unsafe{} usage labels observed in GitHub projects.}
    \label{tbl:labels}
    \begin{subtable}[c]{0.95\textwidth}
    \centering
    \subcaption{An overview of the possibilities for label \lOne~(what is done).}
    \label{tbl:label1}
    \begin{tabular}{rllrr}
        \toprule
        Usage & Description & Code Example & C~\cite{costa2021breaking} & L~\cite{lauinger2020uncovering} \\
        \midrule
        Cast & Implement casts between types & \mintinline{go}{o = (*int32)(unsafe.Pointer(i))}& \covered{} & \covered{}\\
        Definition & Declaration \mintinline{go}{unsafe.Pointer} & \mintinline{go}{var p unsafe.Pointer} & \notcovered{} & \covered{} \\
        Delegate & Pass \unsafe~variable & \mintinline{go}{needPointer(ptr)} & \notcovered{} & \covered{} \\
        Memory-access & Manipulate or reference memory & \mintinline{go}{d := *((*unsafe.Pointer)(ptr))}  &\covered{} & \covered{}\\
        Pointer-arithmetic & Perform arithmetic change of addresses & \mintinline{go}{u := uintptr(unsafe.Pointer(&v[0])) & 3} & \covered{} &\covered{} \\
        Syscall & Use packages for syscalls & \mintinline[breaklines,breakafter=G]{go}{syscall.Syscall(SYS_WRITE, a, uintptr(b), c)} & \covered{} & \covered{}\\
        Unused & Dead code or unused parameters & \mintinline{go}{func A(ptr unsafe.Pointer){}} & \notcovered{} & \covered{}\\
        \bottomrule
    \end{tabular}
    \end{subtable}
    
    \vspace{2mm}
    \begin{subtable}[c]{0.65\textwidth}
    \centering
    \subcaption{An overview of the possibilities for label \lTwo~(underlying purpose of the usage).}
    \label{tbl:label2}
    \begin{tabular}{rlrr}
        \toprule 
        Usage & Description & C~\cite{costa2021breaking} & L~\cite{lauinger2020uncovering} \\ 
        \midrule
        Atomic Operations & Use \mintinline{go}{atomic} package &   \notcovered{} & \covered{}\\
        Avoid Garbage Collection & Prevent free of value &  \notcovered{} & \covered{}\\
        Efficiency & Improve efficiency of program &  \partlycovered{} & \covered{}\\
        Foreign Function Interface & Integrate C code &  \covered{} & \covered{}\\
        Generics & Implement generic functionality &  \notcovered{} & \covered{}\\
        Hide Escape Analysis & Break escape analysis chain &  \notcovered{} & \covered{} \\
        Memory Layout Control & Manage low-level memory &   \partlycovered{} & \covered{}\\
        Reflection & Use or re-implement reflection &  \covered{} & \covered{}\\
        Serialization & Implement marshalling and serialization &   \covered{} & \covered{}\\
        Types & Implement Go type systems (std lib) &  \notcovered{} & \covered{}\\
        Unused & Dead code or unused parameters & \notcovered{} & \covered{} \\
        \bottomrule
    \end{tabular}
    \end{subtable}
    \begin{minipage}{0.28\textwidth}
    \footnotesize
    Note, that we combined the different casts, such as \textit{cast-basic} and \textit{cast-bytes} for brevity for \lOne{} and omit code examples for label \lTwo{} as these are already covered within one of the \lOne{} category examples. 
   \newline 
    \textbf{Legend:}
    \newline
    \underline{C~\cite{costa2021breaking}} and \underline{L~\cite{lauinger2020uncovering}} presents if this usage pattern was discussed by Costa et al.~\cite{costa2021breaking} and Lauinger et al.~\cite{lauinger2020uncovering}, respectively. 
    \newline
    \covered{}: covered,\newline \partlycovered{}: only covered partly,%
    \newline \notcovered{}: not covered.
    \end{minipage} 
    
\end{table*}

\section{Unsafe Usage Patterns}
\label{sec:unsafepatterns}

In this section, we introduce Go's \unsafe{} package along with results~\cite{lauinger2020uncovering,costa2021breaking} on its usages (\autoref{sec:unsafepkg}), the labeled data set used to train our model (\autoref{sec:bckgr:groundtruth}), and one example of an \unsafe~usage (\autoref{sec:exampleusage}). 

\subsection{Unsafe In Go}
\label{sec:unsafepkg}

\textbf{Type Safety and the \unsafe{} Package.}
Go is a statically-typed language.
Like in other type-safe languages, e.g., Java or Rust, the \unsafe{} package provides a way to enable developers to write low-level code and escape type-safety. 

The API consists of five functions and one type~\cite{golang2022unsafe}. 
The type \code{Pointer} represents a pointer type that is more powerful than a "classical" pointer in Go and enables to read and write arbitrary memory. 
The three functions \code{Alignof}, \code{Offsetof}, and \code{Sizeof} provide information about the memory alignment of Go types.
These three functions and one type are discussed in previous studies~\cite{lauinger2020uncovering,costa2021breaking}.
The Go 1.17 release (August 2021) added two new functions to the \unsafe{} package to supposedly simplify the correct usage of the type \code{unsafe.Pointer}. 
As the previous studies did not cover these additions, our discussions of the \unsafe{} package focus on the functions and types introduced before Go 1.17.

By using the \unsafe{} package, developers gain more control over the memory at the cost of potentially introducing vulnerabilities in code, e.g., buffer overflows. 
Besides vulnerabilities, the program may behave differently than expected, e.g., crash. 
In addition, the usage may render the program to be not portable to different systems as well as not being protected by the Go~1 compatibility guideline~\cite{golang2022unsafe}.

\textbf{Usage of \unsafe{} in Go software.}
The \unsafe{} package is used frequently in popular Go projects on GitHub.
Previous empirical studies~\cite{costa2021breaking,lauinger2020uncovering} revealed that
24\% to 38\% of the projects use \unsafe{} within the application code. 
Furthermore, 91\% of the projects use \unsafe{} in transitively imported packages~\cite{lauinger2020uncovering}  with an average depth of $3.08 \pm 1.62$~\cite{lauinger2020uncovering}. 

\subsection{Labels for unsafe.Pointer Usages}
\label{sec:bckgr:groundtruth}

The manual analyses conducted by Costa et al.~\cite{costa2021breaking} and our previous work~\cite{lauinger2020uncovering} reveal several usage patterns for \unsafe{}.
These patterns were observed in a diverse set of applications collected from GitHub and represent usages that occur in-the-wild.
We present these patterns along with code examples in Table~\ref{tbl:labels}. 
The table is divided into two label categories~\cite{lauinger2020uncovering} and 
includes the usage, a description, a code example, and information if the pattern was discussed in previous work. 

The first label category, hereafter \lOne{} (\cref{tbl:label1}), labels what is actually done with the \unsafe{} usage.
One use case for \unsafe{} is to perform casts from arbitrary types to other types, basic types, slices, or \mintinline{go}{unsafe.Pointer} values. 
In \cref{tbl:label1}, we grouped them within the label \textit{cast}. 
As each \unsafe{} usage is labeled, it is possible that a usage "only" declares an \mintinline{go}{unsafe.Pointer} without using it further at this location (\textit{definition}). 
Similarly, to \textit{definition}, it is possible that a usage "only" passes an \unsafe{} variable (\textit{delegate}), e.g., as a parameter. 
The label \textit{memory-access} groups all \unsafe{} usages that manipulate or reference memory.
\textit{Pointer-arithmetic} contains \unsafe{} usages that perform arithmetic changes of addresses, e.g., advancing an array. 
For the interaction with low-level operating system primitives, calls to the \mintinline{go}{syscall} package are necessary, and some functions require \unsafe{} parameters to work correctly (\textit{syscall}). 
Finally, \textit{unused} includes all \unsafe{} usages that are dead code or unused parameters. 

\begin{figure*}
  \begin{subfigure}[t]{1\textwidth}
    \renewcommand{\thesubfigure}{a}
    \begin{minted}[numbers=left, escapeinside=!!, fontsize=\footnotesize]{go}
    // toAddrPointer converts an interface to a pointer that points to the interface data.
    func toAddrPointer(i *interface{}, isptr bool) pointer { !\label{lst:l:toAddrPointer}!
    	if isptr { !\label{lst:l:typebool}!
    		return pointer{p: unsafe.Pointer(uintptr(unsafe.Pointer(i)) + ptrSize)} !\label{lst:l:unsafeusage}!
    	}
    	... 
    }
    \end{minted}
    \caption{A usage of \unsafe{} in a frequently used \textit{protobuf} fork~\cite{protobufpointerunsafe}.}
        \vspace{2mm}

    \end{subfigure}
    \begin{subfigure}{0.48\textwidth}
        \renewcommand{\thesubfigure}{b}
        \begin{minted}[breaklines,numbers=left,firstnumber=last, fontsize=\footnotesize, escapeinside=!!]{go}
func makeMapMarshaler(f *reflect.StructField) (sizer, marshaler) { !\label{lst:l:makeMapMarshaler}!
   ...
   vaddr := toAddrPointer(&vi, valIsPtr) !\label{lst:l:call}!
   ...
}
        \end{minted}
        \caption{One usage of the function with \unsafe{} in the project~\cite{protobuftablemarshal}.}
    \end{subfigure}
    \hfill
    \begin{subfigure}{0.48\textwidth}
    \renewcommand{\thesubfigure}{c}
    \begin{minted}[breaklines,numbers=left,firstnumber=last,fontsize=\footnotesize, escapeinside=!!]{go}
// In pointer_reflect.go we use reflect instead of unsafe to implement the same (but slower) interface.  !\label{lst:l:slowerreflect2}!
type pointer struct { !\label{lst:l:pointerstruct}!
	p unsafe.Pointer
}
    \end{minted}
  \caption{An indication that \unsafe{} is used to improve performance~\cite{protobufpointerunsafe}.}
    \end{subfigure}
\captionof{lstlisting}{An example of an \unsafe{} usage that challenges classification. %
We removed some comments for brevity.}
\label{lst:unsafeclassificationexample}
\end{figure*}
\addtocounter{figure}{-1}

The second label category, hereafter \lTwo{} (\cref{tbl:label2}), focuses on the rationale for the usage.
The package \mintinline{go}{atomic} requires \unsafe{} pointers for some of their functions.
Therefore, developers have to use the \unsafe{} package to interact with the library. 
Go has a garbage collector (GC), and in some cases developers want to prevent that a value is collected by the GC (\textit{avoid garbage collection}) with the help of \unsafe{} usages. 
The label \textit{efficiency} groups usages that aim to improve the time or space of the code. 
Costa et al.~\cite{costa2021breaking} focuses on optimizations due to cast operations. 
While this holds for the majority of usages classified as \textit{efficiency} by us~\cite{lauinger2020uncovering}, we included cases such as \textit{memory-access}.
The label \textit{foreign function interface} (FFI) marks usages that interact with C code, e.g., by calls that expect unsafe pointers. 
During the study, generics were not part of the language. 
Thus, the label \textit{generics} groups \unsafe{} usages where developers implement some generics functionality by themselves. 
The Go compiler has a phase for escape analysis, and in some cases, the developer wants to break the escape analysis chain to improve efficiency~\cite{lauinger2020uncovering}, which is labeled as \textit{hide escape analysis}. 
The label \textit{memory layout control} marks usages that aim to manage the memory. 
The usage pattern by Costa et al.~\cite{costa2021breaking} includes examples for getting the memory address, while our patterns~\cite{lauinger2020uncovering} also include examples for delegation and definition. 
The label \textit{reflection} groups usages that use the \mintinline{go}{reflect} package or implement some reflective functionality.
Usages that (un)marshal or (de)serialize are grouped within the label \textit{serialization}. 
As the labeled data set includes usages of the standard library of Go, some of the \unsafe{} usages implement the type system of Go (\textit{types}). 
Like the \lOne{} label, we group usages that are dead code or unused parameters as \textit{unused}.

\subsection{Example of an \unsafe{} usage}
\label{sec:exampleusage}

Below, we briefly discuss an \unsafe{} usage (\cref{lst:unsafeclassificationexample}).
The function \mintinline{go}{toAddrPointer} (\cref{lst:unsafeclassificationexample}a, Line~\ref{lst:l:toAddrPointer}) casts an empty interface, which can be any type, into a pointer to the data of the interface~\cite{protobufpointerunsafe}. 
Line~\ref{lst:l:unsafeusage} takes the passed interface (\mintinline{go}{i}) to retrieve an \mintinline{go}{unsafe.Pointer} that is cast to a \mintinline{go}{uninptr} and back to a \mintinline{go}{unsafe.Pointer}.
For the conversion back to an \mintinline{go}{unsafe.Pointer} it is necessary to add the offset (\mintinline{go}{ptrSize}). 
The retrieved \mintinline{go}{unsafe.Pointer} is used to initialize the \mintinline{go}{pointer} struct in Line~\ref{lst:l:pointerstruct}, which is returned by the method \mintinline{go}{toAddrPointer}. 
This function is called (\cref{lst:unsafeclassificationexample}b, Line~\ref{lst:l:call}) by the function \mintinline{go}{makeMapMarshaler} (\cref{lst:unsafeclassificationexample}b, Line~\ref{lst:l:makeMapMarshaler}) that marshals a map. 

For \lOne{}, 
we
classified the usage in Line~\ref{lst:l:unsafeusage} as \textit{pointer-arithmetic}. %
In our previous work~\cite{lauinger2020uncovering}, we decided for this label 
because pointer arithmetic was necessary to cast back to \mintinline{go}{unsafe.Pointer}. 
Another possible label is a \textit{cast} due to the conversion. %
The \lTwo{} label for the usage in line~\ref{lst:l:unsafeusage} is \textit{serialization}. 
We decided for this label as
the caller of the function (\cref{lst:unsafeclassificationexample}b, Line~\ref{lst:l:makeMapMarshaler}) marshals a message. 
Nonetheless, it could be argued that \lTwo{} for the usage is \textit{efficiency}, because in the global context of the program, \unsafe{} is used to improve the efficiency in comparison to the reflection-based implementation as indicated in Line~\ref{lst:l:slowerreflect2}. 

This example illustrates two possible refactorings to harden or even avoid the \unsafe{} usage.
First, in Line~\ref{lst:l:unsafeusage}, one can use the function \mintinline{go}{Add}, which was newly introduced in Go and can be used to wrap the pointer arithmetic. 
Second, it may be possible to refactor the usage of \unsafe{} entirely by evaluating currently existing marshaling libraries. 
For both cases, it is essential to identify and classify \unsafe{} usages. 
\section{Unsafe Code Classification}%
\label{sec:design}

\begin{figure*}
\centering
\includegraphics[width=0.7\textwidth]{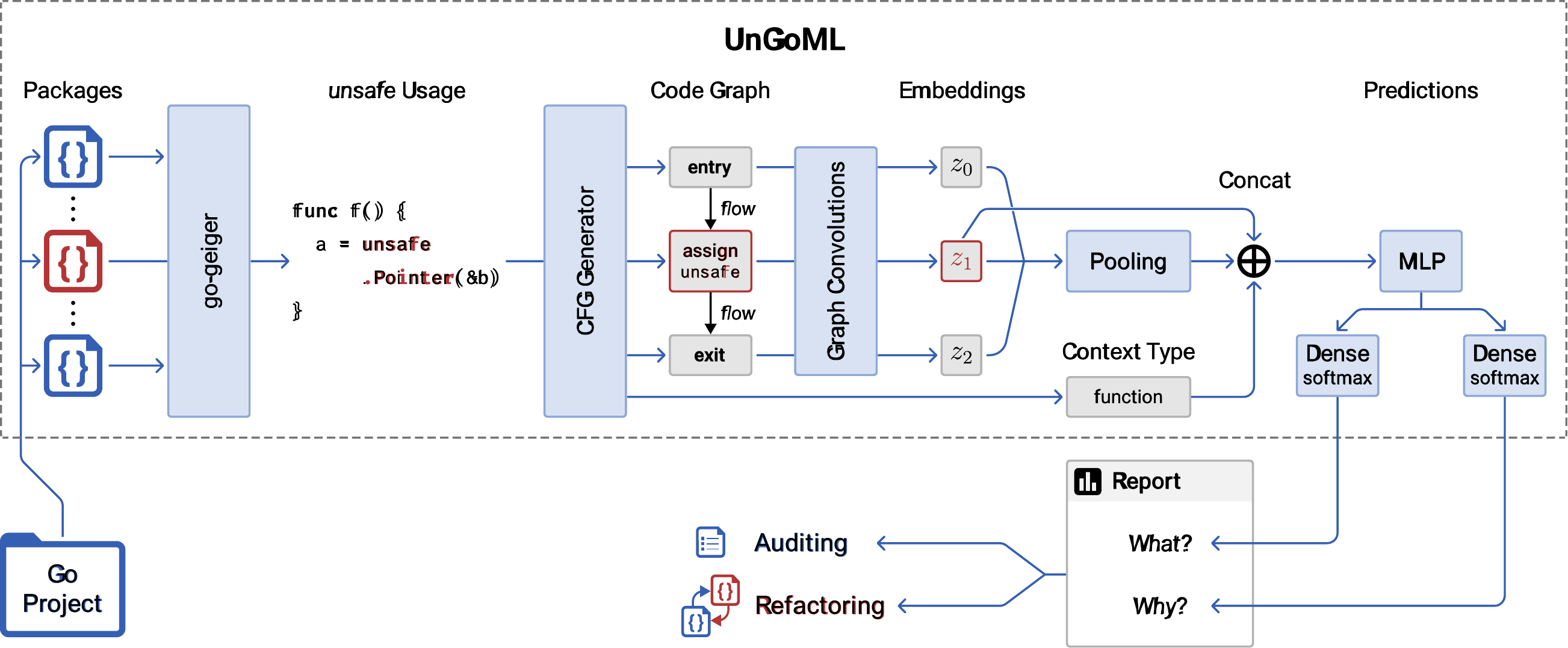}
\caption{High-Level Design of \ourtool.}
\label{fig:highleveldesignungoml}
\end{figure*}

This section presents our approach to automatically classify a given \unsafe~usage, thereby answering the two questions ``What is happening?'' and ``For what purpose?''.
Figure \ref{fig:highleveldesignungoml} shows a high-level overview of the composition of \ourtool. 
\Cref{sec:design:representation} focuses on the representation of an \unsafe~usage, \cref{sec:design:model} then describes how this representation is used to classify the usage.

\subsection{Code Representation}%
\label{sec:design:representation}

We represent \unsafe{} usages as enriched \acfp{cfg}, which encode information about the usages and their surroundings.
Thus, our approach follows a recent trend observed for vulnerability detection, that programs are encoded into a \ac{cfg} variant~\cite{sonnekalb_deep_2022}.
We developed our \ac{cfg} representation by investigating information that is relevant to our problem and can be easily derived.
Given \unsafe{} usages as pointers to lines of Go code, their graph representation contains the control-flow structure of their surrounding context. 
Possible contexts of an usage are either the body of the function, the type declaration, or the global variable definition where the usage occurs.
In cases where the \unsafe{} usage context is a type declaration or a global variable definition, there is no control-flow structure, and the context is represented as a single statement vertex of type \texttt{declaration}. 

There are two types of vertices in the \ac{cfg} representation: \textit{Statement} vertices and \textit{variable} vertices.
\emph{Statement vertices} correspond to Go statements; they are connected via 
\texttt{flow} and \texttt{alt-flow} edges, which indicate possible execution paths -- the latter represent the control-flow from a branching statement to its successor if the branch condition is not satisfied; \texttt{flow} edges represent all other control-flow relations.
\emph{Variable vertices} correspond to the \emph{named} memory locations referenced by statement vertices; this includes stack and heap variables, as well as struct fields and function pointers.
Edges from statement vertices to variable vertices represent different types of memory accesses, namely \texttt{decl} edges for variable declarations, \texttt{use}/\texttt{dir-use} for reads, \texttt{update}/\texttt{assign} for writes, and \texttt{call} for function pointer calls.

There are two types of read access edges to distinguish between tail and non-tail positions in pointer dereference chains of the form \texttt{x = \textcolor{t_blue}{s.f1.f2}\textcolor{t_red}{.f3}}; the three non-tail dereferences would be represented by \textcolor{t_blue}{\texttt{use}} edges, while the final dereference of the field \textcolor{t_red}{\texttt{f3}} would be represented by a \textcolor{t_red}{\texttt{dir-use}} edge.
Analogous, there are two types of write edges to distinguish between direct (``real'') writes to memory and indirect variable modifications; 
a statement of the form \texttt{\textcolor{t_blue}{s.f1.f2}\textcolor{t_red}{.f3} = x} would have three \textcolor{t_blue}{\texttt{update}} edges to the non-tail dereferences and one \textcolor{t_red}{\texttt{assign}} edge to the field \textcolor{t_red}{\texttt{f3}}.

Lastly, there are \texttt{contains} edges from a variable vertex $v_a$ to a variable vertex $v_b$.
They indicate the existence of a pointer to the memory location of $v_b$ at $v_a$; for the previous example \texttt{s.f1.f2.f3}, there would be three \texttt{contains} edges in our graph: $\texttt{s}\to\texttt{f1}$, $\texttt{f1}\to\texttt{f2}$ and $\texttt{f2}\to\texttt{f3}$.

\begin{figure*}[t]
    \centering
    \includegraphics[width=0.85\linewidth]{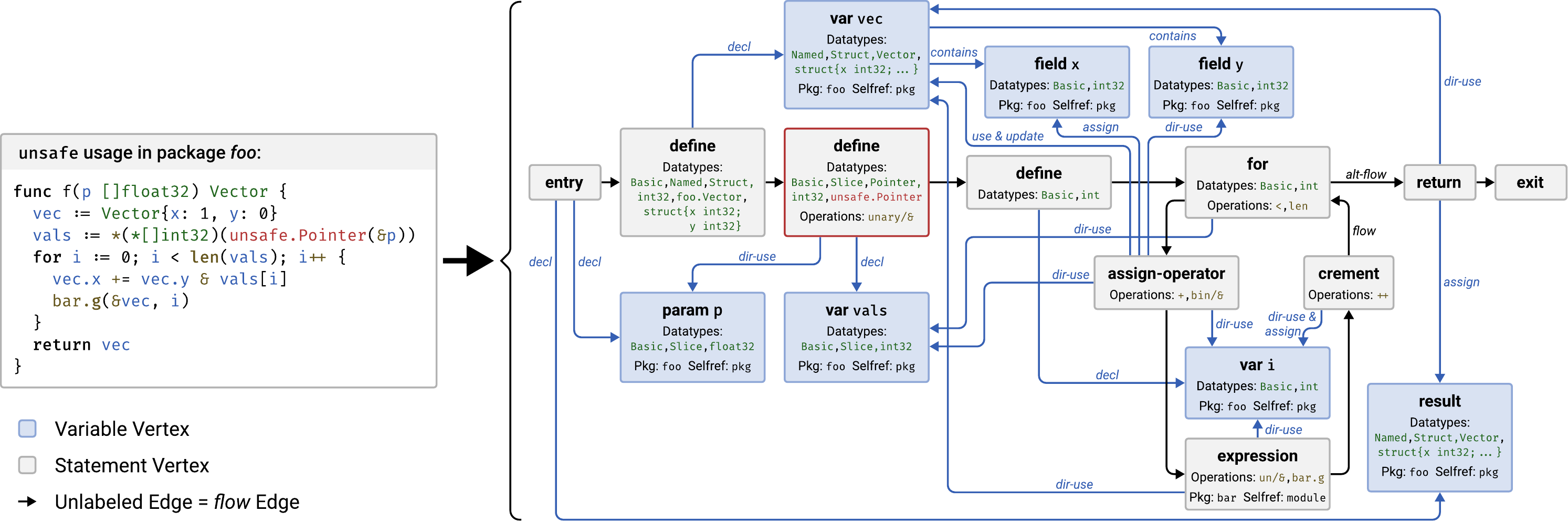}
    \caption{
        An exemplary code snippet containing \unsafe{} and its corresponding enriched \ac{cfg} representation. We highlight the "enriched" part of our \ac{cfg} with the blue nodes and edges, and the \unsafe{} usage with red.
    }\label{fig:design:example}
\end{figure*}

\subsubsection{Mapping Go Code to Vertices}%
\label{sec:design:representation:vertices}
Most \ac{cfg} vertices directly correspond to a single Go statement or variable/field.
However, there are two exceptions to this direct correspondence.
First, we add a \texttt{flow} edge between the pseudo-statement vertex \texttt{entry} and the first Go statement and one for all terminal statements, such as \texttt{return}, to \texttt{exit}.
Further, we model function parameters, return values, and receivers as special variable vertices that are declared by the \texttt{entry} vertex.
\texttt{return} statements are treated as assignments to the return variables, with \texttt{assign} edges being added between both.
Second, we split branch statements that combine a branch condition expression with another statement, as in \mintinline{go}{for i := 0; i < 6; i++ {...}}, into separate statement vertices. 
\Cref{fig:design:example} illustrates the \ac{cfg} vertices that are generated for a simple function. 

\subsubsection{Label-based Vertex Representation}%
\label{sec:design:representation:labels}
We encode the individual statements and variables by assigning labels to each vertex.
Statement vertices can have the following labels:

    \paragraph*{\textbf{Statement Types}}
        Each statement vertex has exactly one label representing its 
        type, e.g., \texttt{if}, \texttt{switch}, \texttt{for}, \texttt{return}, \texttt{assign}, or \texttt{declare}.
    \paragraph*{\textbf{Data Types}}
        The data types that are instantiated in a statement via \mintinline{go}{make}, \mintinline{go}{new}, literal expressions, or casts are represented using distinct labels.
        For composite types, we also include additional labels for encoding the contained basic types (e.g., \texttt{bool} or \texttt{float32}) and the used composition structures (e.g., \texttt{Struct} or \texttt{Slice}). 
        The instantiation of the type \mintinline{go}{map[string]**[]int} inside a statement would be represented by six labels: \texttt{Map}, \texttt{string}, \texttt{Pointer}, \texttt{Slice}, \texttt{int}, and one label for the complete composite type.
    \paragraph*{\textbf{Operators}}
        Go operators occurring in a statement are represented by corresponding operator labels, e.g., \texttt{binary/+}, \texttt{binary/\allowbreak{}==}, \texttt{unary/\&}, or \texttt{unary/-}.
    \paragraph*{\textbf{Functions}}
        Both the built-in Go functions (e.g., \texttt{len} or \texttt{append}) and all other regular package functions (e.g., \texttt{fmt.Errorf} or \texttt{golang.org/\allowbreak{}x/sys/unix.Syscall}) called within a statement are represented by distinct labels.
    \paragraph*{\textbf{Packages}}
        In addition to the function labels, the origin packages of all called functions in a statement are added as labels, e.g., \texttt{fmt} or \texttt{golang.org/x/sys/unix}.
    \paragraph*{\textbf{Self-references}}
        If a statement in the body of a function $f$ contains a recursive call to $f$, the \texttt{selfref/function} label is added to that statement.
        If a statement occurring in a Go module $m$ and a package $p$ contains a call to any function from $m$ and/or $p$, the \texttt{selfref/module} and/or \texttt{selfref/package} labels are added.
        Lastly, recursive declaration statements, e.g., recursive structs, are labeled with the \texttt{selfref/type} label.
\noindent

Note that this label-based representation does not preserve the syntactic structure within statements; both statements \mintinline{go}{x=f(a+b)*g(f(c))} and \mintinline{go}{x=g(a*f(b)+c)} will be represented by the same set of vertex labels.
We also experimented with a more fine-grained representation that encodes the \acp{ast} of statements.
However, this representation did not improve the classification accuracy, while increasing graph sizes and therefore slowing down model training; hence, we chose to discard the \ac{ast} structure of statements.

Variable vertices are labeled using the following categories:
    \paragraph*{\textbf{Variable Types}}
        We model function parameters, results, and receivers as variable nodes. 
        To distinguish those special variables from regular Go variables, we add a \texttt{param}, \texttt{result}, or \texttt{receiver} label to them.
    \paragraph*{\textbf{Variable Names}}
        The name of a variable is added as an additional label to each variable vertex,
        such as \texttt{i} for iterators or \texttt{err} for errors. 
    \paragraph*{\textbf{Datatypes}}
        We represent the data type of each variable by a set of labels analogously as for statement vertices. 
    \paragraph*{\textbf{Packages}}
        For each variable, we add the name of the package in which the variable is defined as an additional label. 
    \paragraph*{\textbf{Self-references}}
        If a variable is defined in the same 
        module and/or package as the context of the considered \unsafe{} usage, a \texttt{selfref/module} and/or \texttt{selfref/package} is added.
        If the context of an \unsafe{} usage is a global variable definition, the variable vertex corresponding to the defined global variable gets the \texttt{selfref/variable} label.
\noindent
The \ac{cfg} in \cref{fig:design:example} illustrates how different label types are used. 

\subsubsection{Mapping Vertex Labels to Feature Vectors}%
\label{sec:design:representation:features}

One common assumption of \ac{ml} algorithms is that their input is represented as one or multiple numerical feature vectors of fixed dimensionality.
This is a common requirement that also holds for other machine learning models besides \acp{gnn}, such as \codebert~\cite{feng2020codebert}.
The labeling scheme described in the last section does not fulfill this assumption; the individual labels are not numerical nor is their number fixed, since there is a potentially infinite number of data types, functions, packages, and variable names.

We address this issue by restricting the set of allowed labels to the most common ones in the training data.
More precisely, for each of the label categories 
the top-$k$ most frequent labels within that category are selected.
For finite label categories, $k$ is chosen such that all possible labels within those categories are selected.
For the four infinite label categories \textsc{Datatypes}, \textsc{Functions}, \textsc{Packages} and \textsc{Variable Names}, we chose a fixed cutoff of $k = 127$ as we did not observe any improvements by using a larger cutoff in our experiments.
If a vertex has an uncommon label that is not part of the per-category top-$k$ selection, that label is replaced by a fallback `\texttt{other}' label for its category.
For example, variable names such as \texttt{err} for errors and \texttt{i} for iterators are encoded, while uncommon variable names are grouped within the label `\texttt{other}'. 
This approach reduces the number of considered labels for each infinite category to $k + 1 = 128$.
Combining the labels of all categories, we obtain a total of $n \coloneqq 4 \cdot (k+1) + n_{\mathit{finite}} = 594$ possible vertex labels $\mathcal{L} = {\{l_i\}}_{i=1}^{n}$, where $n_{\mathit{finite}}$ is the number of labels in the finite label categories.

Through this reduction of the permitted labels, the label set $L_v \subseteq \mathcal{L}$ of a vertex $v$ can be encoded as a binary feature vector $x_v \in {\{0, 1\}}^n$, with $x_v{[i]} \coloneqq \mathbbm{1}[l_i \in L_v]$.
Using this encoding strategy, an enriched \ac{cfg} becomes a directed multigraph with binary vertex feature vectors and nine types of edges.

\subsection{Model Architecture}%
\label{sec:design:model}

Our \unsafe~usage classification approach is based on a family of \ac{ml} models called \acfp{gnn}.
Over the recent years, \acp{gnn} have been successfully applied to various graph learning tasks, including graph classification, and are becoming the standard for software engineering classification tasks such as vulnerability detection ~\cite{chakraborty2021deep, sonnekalb_deep_2022, zhou2019devign, cheng2021deepwukong}.
\Acs{gnn}-based graph classifiers are a family of differentiable models, which take graphs with vertex, and depending on the \ac{gnn} variant, edge feature vectors as input and output a vector encoding the predicted class probabilities.
Below, we briefly introduce \acp{gnn} and subsequently 
present how we use them to solve the \unsafe~usage classification problem.

\subsubsection{Introduction to \acsp*{gnn}}%
\label{sec:design:model:gnn}

A \ac{gnn} for graph-level prediction tasks, such as graph classification, typically consists of a sequence of so-called \emph{graph convolution} layers, followed by a \emph{graph pooling} layer.
Generally speaking, a graph convolution operator takes a set of vertex feature vectors ${\{x_i \in \mathbb{R}^{d} \}}_{i=1}^n$ as input\footnote{There are convolution approaches that also consider edge feature vectors. 
For simplicity, they will not be covered here.} and aggregates the feature vector $x_i$ of each vertex $v_i$ with the feature vectors of other vertices $v_j$ that are related to $v_i$ by some structural characteristic in the graph.
The result of each convolution is a set of 
convolved feature vectors ${\{z_i \in \mathbb{R}^{d'} \}}_{i=1}^n$.
After applying one or more graph convolutions to the feature vectors, a final graph-level vector representation $z_G \in \mathbb{R}^{d'}$ is obtained by combining the convolved vectors ${\{z_i \in \mathbb{R}^{d'} \}}_{i=1}^n$ via a pooling layer.

The simplest possible graph convolution ignores all 
structural information and treats a graph as a set of vertices.
It aggregates each vertex feature $x_i$ only with itself; 
this convolution is described by $z_i = f(x_i)$,
$f: \mathbb{R}^d \to \mathbb{R}^{d'}$.
If $f$ is chosen to be a \ac{mlp}, one obtains a so-called \emph{DeepSets}~\cite{Zaheer2017} model.
Next, we look at \acp{gnn} that do utilize graph structure.

Most graph convolution approaches are based on the principle of aggregating vertices that are spatially related, typically by being direct neighbors.
The so-called \ac{gcn} convolution~\cite{Kipf2017}, for example, updates the feature vector of each vertex by computing the mean of the features of its neighbors.
Xu et al.~\cite{Xu2018} show that this approach has a limited \emph{discriminative power}, i.e., it cannot distinguish some classes of non-isomorphic graphs.
To address this limitation, they propose the more powerful
\ac{gin} architecture.

Recently, multiple approaches going beyond the 
discriminative power of \ac{gin}
have been proposed~\cite{Damke2020,Bouritsas2020,Maron2019}.
The so-called \acs{wl2gnn} architecture~\cite{Damke2020}, for example, is based on the 2-dimensional (Folklore) \ac{wl} graph isomorphism test~\cite{Cai1992}; it is provably more powerful than \ac{gin}.

\subsubsection{Unsafe Usage Classification}%
\label{sec:design:model:architecture}

The architecture of our \unsafe{} usage classification model follows the standard \ac{gnn} structure for graph-level prediction tasks, i.e., it consists of a sequence of graph convolution layers, followed by a graph pooling layer, which produces a vector embedding for a given \ac{cfg}.
This embedding is a summary of the encoded information and reduces the size of the feature map.
This pooled embedding vector is then concatenated, to compensate for the loss introduced by the pooling layer, with two additional vectors that encode the following information:

    \paragraph*{\textbf{Usage Location}}
    A Go function may contain multiple \unsafe{} usages with different \lOne{} and \lTwo{} labels.
        However, in the pooled graph embedding vector, all statements that contain \unsafe{} are merged and can no longer be distinguished.
        We address this issue by concatenating the pooled vector with the convolved feature vector of the vertex representing the Go statement, which contains the \unsafe{} usage that should be labeled.
    \paragraph*{\textbf{Context Type}}
   Additionally, we append a one-hot context type vector, which encodes whether the \unsafe{} usage occurs in a function declaration, a global variable, or a type definition.
        Since global variable and type definitions are both represented as single \texttt{declaration} statement vertices, they generally cannot be distinguished from each other and from functions without parameters and return types whose body only contains a single variable declaration.
        The context type vector allows the model to distinguish these cases.
\noindent
Finally, the concatenated embedding vector is fed into a \ac{mlp}, followed by two parallel fully-connected layers with softmax activations, which produce two label probability distributions for the \lOne{} and \lTwo{} labels.
We present this architecture in \cref{fig:highleveldesignungoml} and mark the \unsafe{} usage with red. The concatenated embedding vector is represented by the edge after the concatenation icon in the right half of the Figure.

For the evaluation in \cref{sec:eval}, we created four variants of this architecture using the following types of graph convolutions:
DeepSets~\cite{Zaheer2017}, \ac{gin}~\cite{Xu2018} and \acs{wl2gnn}~\cite{Damke2020}.
To determine the importance of the context in which an \unsafe{} usage occurs, we additionally built a simple baseline \ac{mlp} model, which only gets the feature vector of the vertex representing the statement
to be classified.

We train our models with the Adam optimizer~\cite{Kingma2015} and the sum of the cross-entropies for both labels as the loss function; i.e.,\ the optimization target is to maximize the predicted probabilities of the correct labels for each usage.
Modern neural network classifiers often tend to be overly confident, i.e., the model's accuracy is lower than the average probability it assigns to its top-1 predictions~\cite{Guo2017}.
To address this issue, we calibrate the predicted probabilities for \lOne{} and \lTwo{} independently via temperature scaling~\cite{Platt1999,Guo2017}.

We do not only output the \lOne{} and \lTwo{} labels with the highest predicted probability but instead output a set of labels for each label category.
The rationale is to account for the fact that correct labels for a given \unsafe{} usage are not always obvious, even to a human expert.
The predicted sets can serve as a preselection that assists the user in determining the most plausible \lOne{} and \lTwo{} labels.
Each prediction set is created by selecting the top-$k$ classes until the probability mass of the set exceeds a certain threshold.
The threshold is chosen via \emph{inductive conformal prediction} using the so-called \emph{\acl{giq} nonconformity score}~\cite{Romano2020}.
This approach provably guarantees that the predicted \lOne{} and \lTwo{} sets each contains the respective true label for a given usage with a probability of at least $1 - \varepsilon$, where $\varepsilon \in {[0,1]}$ is a significance level that can be freely chosen; we use the common default $\varepsilon = 0.1$.
In addition to this validity guarantee, the sizes of the conformal prediction sets are adaptive; if the model is uncertain about the label for a given usage, it will produce a large prediction set. Likewise, a small set is produced if the model is certain about the true label.

\section{Implementation and Evaluation}
\label{sec:eval}

We implemented our approach as a self-contained tool, \ourtool{}, which provides functionality to both identify and quantify \unsafe{} snippets (\geiger{}~\cite{lauinger2020uncovering}) along with the classification for \unsafe{} usages (\cref{fig:highleveldesignungoml}).
\ourtool{} expects a Go project, e.g., local or a GitHub URL, as input and returns a report of \unsafe{} usages within the Go project, including the prediction of \lOne{} and \lTwo. 
To generate this report, \ourtool{} runs \geiger{} upon the project. 
Afterward, we pass all identified \unsafe{} usages to the classifier, which predicts for each snippet \lOne{} and \lTwo{} labels. 
These predictions are combined with the results of \geiger{} to generate a report of all \unsafe{} usages within the given project.

In \autoref{sec:eval:dataset} we describe the studied dataset, followed by our experimental setup in \autoref{sec:eval:setup}, and 
answer the following three research questions in \autoref{sec:eval:results}

{\setlength\leftmargin{1pc}\begin{description}
    \item[RQ1:] \textbf{What is the impact of the context in predicting \lOne{} and \lTwo{} of \unsafe{} usages?}
    The \ac{gnn} models get the context of a usage as input, while the \ac{mlp} baseline only considers a single vertex.
    \item[RQ2:] \textbf{What is the impact of control- and data-flow for classifying \unsafe{} usages?}
    Control- and data-flows are commonly used, e.g., in static analyses, to reason about code, and are effective for deep learning vulnerability detection 
    \cite{sonnekalb_deep_2022, chakraborty2021deep, zhou2019devign}. 
    \item[RQ3:] \textbf{How relevant are different vertex features for the classification of \unsafe{} usages?}
    The vertex labels in our enriched \ac{cfg} representation encode various aspects of a code snippet.
    We want to determine the importance of the different vertex label categories, e.g.,\ \textsc{Datatypes} or \textsc{Functions}. 
\end{description}}

We also present the predictions for \cref{lst:unsafeclassificationexample} in \autoref{sec:eval:examplepred}, and discuss in \autoref{sec:eval:applications} how \ourtool{} can be used in practice.

\subsection{Studied Dataset}
\label{sec:eval:dataset}

We chose our dataset~\cite{lauinger2020uncovering}
over the one of Costa et al.~\cite{costa2021breaking} for the following reasons.
First, with 1,400 entries, it is much larger than the 270 labeled entries of Costa et al.
~\cite{costa2021breaking}.
Second, the labels are provided in two dimensions instead of one, thus providing a more fine-granular representation of the \unsafe{} usage. 
Further, our comparison of the usage patterns~(cf.~\cref{tbl:labels}) confirms that we identified more diverse \unsafe{} usage patterns. 
Third, we 
~\cite{lauinger2020uncovering} labeled \unsafe{} usages on a statement-level instead of a file-level. 
Thus, we can predict usages more fine-granular. 
Fourth, the analysis of Costa et al.
~\cite{costa2021breaking} focuses only on \unsafe{} usages within the application code without considering dependencies.
Thus, it ignores a common source of many \unsafe{} usages~\cite{lauinger2020uncovering} and vulnerabilities~\cite{pashchenko_vuln4real_2022, zimmermann_small_2019}. 
To train our model, we used the dataset as-is. 
The labels are derived from the 10 projects of the top-500 starred GitHub projects with the most \unsafe{} usages. 
For more details about this data set, we refer to our paper 
~\cite{lauinger2020uncovering}. 

\subsection{Experimental Setup}%
\label{sec:eval:setup}

To train and evaluate different model architectures, we use an existing manual labeled data set of \unsafe{} usages~\cite{lauinger2020uncovering} (more details \cref{sec:bckgr:groundtruth}). 
We randomly split the dataset into ten stratified folds of equal size~\cite{hall2012state}, preserving the joint distribution of the \lOne{} and \lTwo{} label dimensions in each fold.
Each model is independently tuned, trained, and evaluated ten times using each bin once as test data and in the other iterations as training/validation data. All following results are averages over the ten iterations.

In each iteration, we further subdivide the training/validation data into a 90\% training split and a 10\% validation holdout split.
As mentioned in \cref{sec:design:model:architecture}, we use the Adam optimizer~\cite{Kingma2015} to minimize the sum of the cross-entropy losses for both labels.
The learning rate is fixed at $0.001$.
To tune the hyperparameters of each model, we use Hyperband~\cite{Li2018a} with a reduction factor of $3$ and a maximum epoch count of $200$.
As the hyperparameter optimization objective, we use the joint top-1 accuracy on the validation split, i.e., the proportion of validation instances for which the \lOne{} and \lTwo{} labels with the highest predicted probability are \emph{both} correct.
The explored hyperparameter space is shown in \cref{tbl:eval:params}.
The activations and widths of the convolution and \ac{mlp} layers are tuned independently.
The \texttt{softmax} pooling treats one vertex feature dimension as a weight distribution logic to compute the weighted mean of the remaining dimensions.
Batch normalization is only applied after convolutions, dropout only between \ac{mlp} layers.
\begin{table}[t]
	\caption{
        Explored hyperparameter space.
    }\label{tbl:eval:params}
	\centering
 	{\setlength{\tabcolsep}{5pt}\renewcommand{\arraystretch}{0.85}%
	\begin{tabular}{rl}
	     \textbf{Parameter} & \textbf{Values} \\ 
	     \toprule
	     Depths & Conv.~$\in \{ 2, \dots, 6 \}$, \acs{mlp} $\in \{ 1, 2, 3 \}$ \\
	     Activations & $\{ \texttt{relu}, \texttt{sigmoid}, \texttt{tanh}, \texttt{elu} \}$ \\ 
	     Layer Widths & $\{ n \in \mathbb{N} \mid {n \pmod{32} \equiv 0} \land n \leq 512 \}$ \\
	     Pooling & $\{ \texttt{sum}, \texttt{mean}, \texttt{max}, \texttt{min}, \texttt{softmax} \}$ \\
	     Regularization & Batch Norm {\normalfont$\in \{\textrm{yes}, \textrm{no}\}$}, Dropout $\in \{ 0, \frac{1}{2} \}$ \\
	     \bottomrule
	\end{tabular}}
\end{table}

After hyperparameter selection, the model is trained three times to estimate the performance variance caused by random weight initialization. %
For each training repeat, we use a maximum epoch limit of 1,000 in combination with early stopping if the validation loss does not decrease for $100$ epochs.
We use the validation split as the calibration data for temperature scaling~\cite{Platt1999,Guo2017} and conformal prediction.
To evaluate each trained and calibrated model, we use the top-1, top-3, and conformal set accuracy, plus the mean conformal set size on the test data.
We aggregate the results for the three repeats by computing the mean and standard deviation for each metric.
Those results for each of the ten outer iterations are further aggregated by computing the expected value and standard deviation of their mean.

\subsection{Model Comparison and Ablation Study}%
\label{sec:eval:results}

\begin{table}[t]
	\caption{
        Unsafe usage combined mean test classification accuracies (in \%) and the set sizes for conformal set prediction. 
    }\label{tbl:eval:resultssmall}
	\centering
	{\sisetup{
        round-mode = places,
        round-precision = 1,
        propagate-math-font = true,
        evaluate-expression
    }\setlength{\tabcolsep}{5pt}\renewcommand{\arraystretch}{0.85}

\csvreader[
		column count=364,
		tabular={c r | r | rrr},
		separator=comma,
		table head={%
			& %
			\multicolumn{1}{c}{\textbf{Top-1 }} &%
			\multicolumn{1}{c}{\textbf{Top-3 }} &%
			\multicolumn{3}{c}{\textbf{Conformal Set }} \\ 
			& %
			\multicolumn{1}{c}{Acc.} &
 			\multicolumn{1}{c}{Acc.} &
 			\multicolumn{1}{c}{\lOne{} Size} &
 			\multicolumn{1}{c}{\lTwo{} Size} &
 			\multicolumn{1}{c}{Acc.} %
			\\\toprule%
		},
		table foot=\bottomrule,
		late after line=\ifthenelse{\equal{\model}{MLP}}{\\\midrule}{\\},
		head to column names,
		filter={\equal{\limitId}{all}}
		]{data/results.csv}{}{%
		\textbf{\model} &%
		{\evalsize{\testAccuracyBestModel}{\testAccuracyBestLimitId}{\testAccuracyWorstLimitId}{100*\testAccuracyMean}{100*\testAccuracyStd}} &%
		{\evalperc{\testAccuracyTopIIIBestModel}{\testAccuracyTopIIIBestLimitId}{\testAccuracyTopIIIWorstLimitId}{100*\testAccuracyTopIIIMean}{100*\testAccuracyTopIIIStd}} &%
		{\evalsize{\testLabelIMeanSizeConfpBBestModel}{\testLabelIMeanSizeConfpBBestLimitId}{\testLabelIMeanSizeConfpBWorstLimitId}{\testLabelIMeanSizeConfpBMean}{\testLabelIMeanSizeConfpBStd}} &%
		{\evalsize{\testLabelIIMeanSizeConfpBBestModel}{\testLabelIIMeanSizeConfpBBestLimitId}{\testLabelIIMeanSizeConfpBWorstLimitId}{\testLabelIIMeanSizeConfpBMean}{\testLabelIIMeanSizeConfpBStd}} & %
		{\evalperc{\testAccuracyConfpBBestModel}{\testAccuracyConfpBBestLimitId}{\testAccuracyConfpBWorstLimitId}{100*\testAccuracyConfpBMean}{100*\testAccuracyConfpBStd}} %
	}}
\end{table}

\Cref{tbl:eval:resultssmall} presents the aggregated results we obtained for the evaluated models trained with all feature dimensions (\cref{sec:design:representation:features}).
We compare the accuracies obtained for all models relative to a majority classifier that predicts the distribution of \lOne{} and \lTwo{} in its training data. 
Overall, we find that all four types of models significantly outperform the baseline majority classifier by at least 45\%, indicating that they successfully learn generalizable correlations between \ac{cfg} properties and the target labels.
Also, all context-aware models perform similarly and have consistently higher accuracies than the MLP models in top-1~(6\%) and top-3~(4\%).

\begin{rqbox}{}{}
The three context-aware models significantly outperform the single vertex \ac{mlp} model, showing that the context is important for \unsafe{} classification. 
\end{rqbox}

The DeepSets model achieves essentially the same mean results as \ac{wl2gnn} without considering control-flow structure and variable usage dependencies. 
Note, that only \ac{wl2gnn} considers edge labels, while \ac{gin} treats all edges equally.
This architecture generally performs slightly worse than the other two context-aware models. 

\begin{rqbox}{}{}
For \unsafe{} classification, the impact of control- and data-flow is negligible, as DeepSets achieves essentially the same accuracies as \ac{wl2gnn} and \ac{gin}.
\end{rqbox}

\begin{figure}
    \centering
    \includegraphics[width=0.45\textwidth]{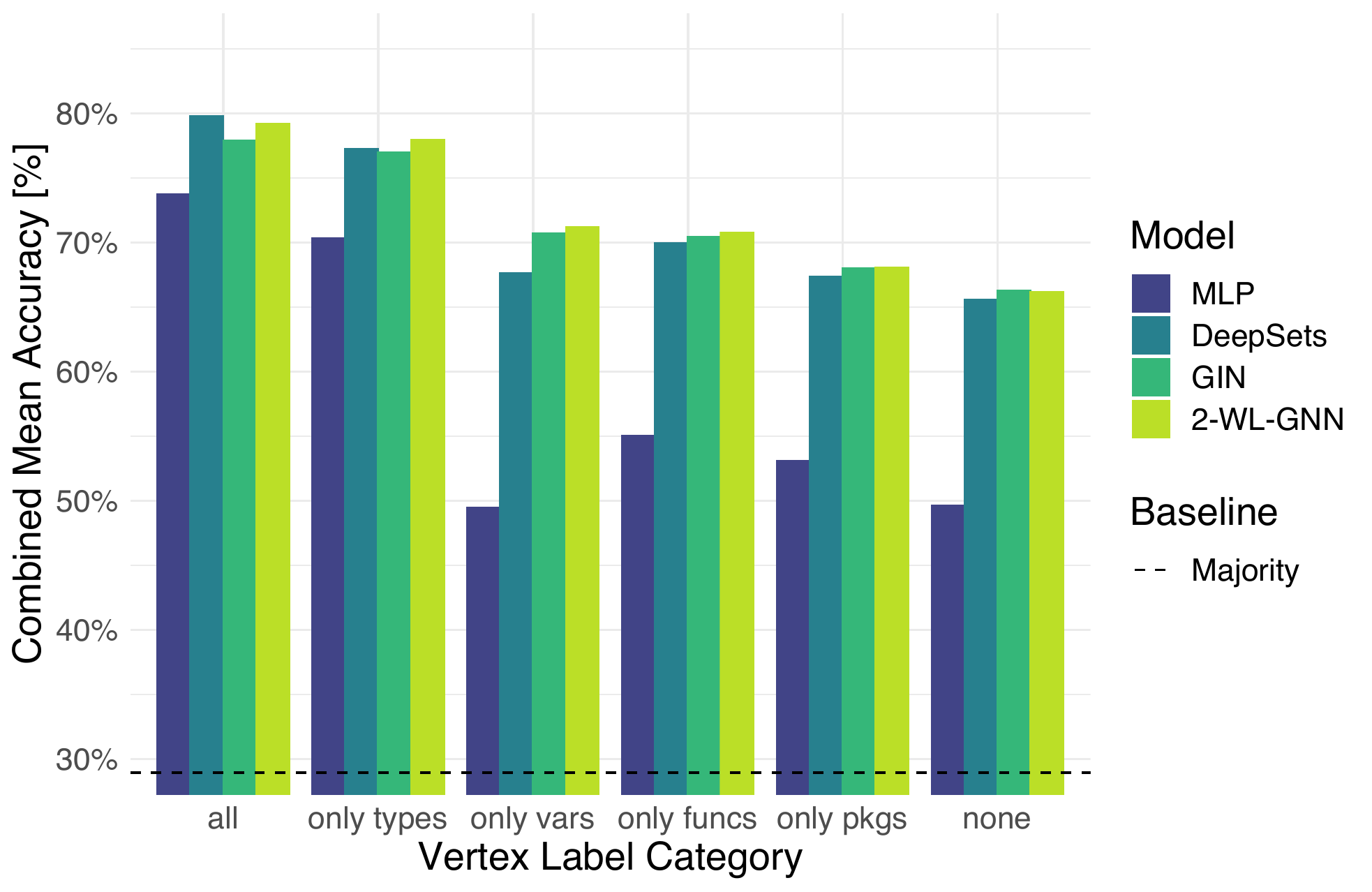}
    \caption{Combined Top-1 Prediction Accuracy for different vertex feature sets. \textsc{all} shows the results for \acp{cfg} with the full set of vertex features. For the remaining, all (\textsc{none}) or all but one infinite vertex label category
    are removed.}
    \label{fig:accuracyvertexcats}
\end{figure}

\Cref{fig:accuracyvertexcats} shows the aggregated results we obtained for the evaluated models when trained with different subsets of the feature dimensions (\cref{sec:design:representation:features}) and models.
The block \textsc{all} contains the results for the models trained with the complete set of features.
In the \textsc{none} block, all variable vertices and all vertex labels, except those from the categories \textsc{Statement Types} and \textsc{Self-references} are omitted.
In the other blocks, the \textsc{none} block is extended by exactly one infinite vertex label category. 
The block \textsc{only vars} includes the variable vertices, 
\textsc{only types} adds the \textsc{Datatypes} labels, and \textsc{only funcs} enters the \textsc{Functions} and \textsc{Operator} labels. 
Lastly, \textsc{only pkgs} includes the \textsc{Packages} labels.
As in \cref{tbl:eval:resultssmall}, we added the majority classifier as a baseline. 

Comparing different feature subsets, we find that the \textsc{Datatypes} labels are the most important -- the block \textsc{only types} comes closest to the models trained with the complete feature set.
This is plausible since the \textsc{Datatypes} label category contains, among others, the \texttt{unsafe.Pointer} and \texttt{uintptr} labels.
The other feature subsets are generally much less informative for classifying \unsafe{} usages.

\begin{rqbox}{}{}
    The most important features for the \unsafe~classification are the \textsc{datatypes}. 
\end{rqbox}

\def\labelIwidth{0.24\linewidth}
\def\labelIIwidth{\dimexpr(\labelIwidth * 122 / 129)}
\def\confHeadShift{0.253\dimexpr\labelIwidth} %
\def\confHeadShiftII{0.212\dimexpr\labelIIwidth} %
\newcommand{\confusionMatrix}[2]{\raisebox{-\height+0.15cm}{\includegraphics[width=#1]{gfx/confusion_matrices/#2.pdf}}}
\newcommand{\cooccMatrix}[2]{\raisebox{-0.5\height}{\includegraphics[width=#1]{gfx/cooccurrence_matrices/#2.pdf}}}
\begin{figure*}[ht]
    \centering
    \setlength{\tabcolsep}{3pt}
    \footnotesize
	\begin{tabular}{ccccc}
	    \hspace*{\confHeadShift}\textbf{MLP} \lOne &%
	    \hspace*{\confHeadShift}\textbf{2-WL-GNN} \lOne &%
	    \hspace*{\confHeadShiftII}\textbf{MLP} \lTwo &%
	    \hspace*{\confHeadShiftII}\textbf{2-WL-GNN} \lTwo %
	    \\
	    \confusionMatrix{\labelIwidth}{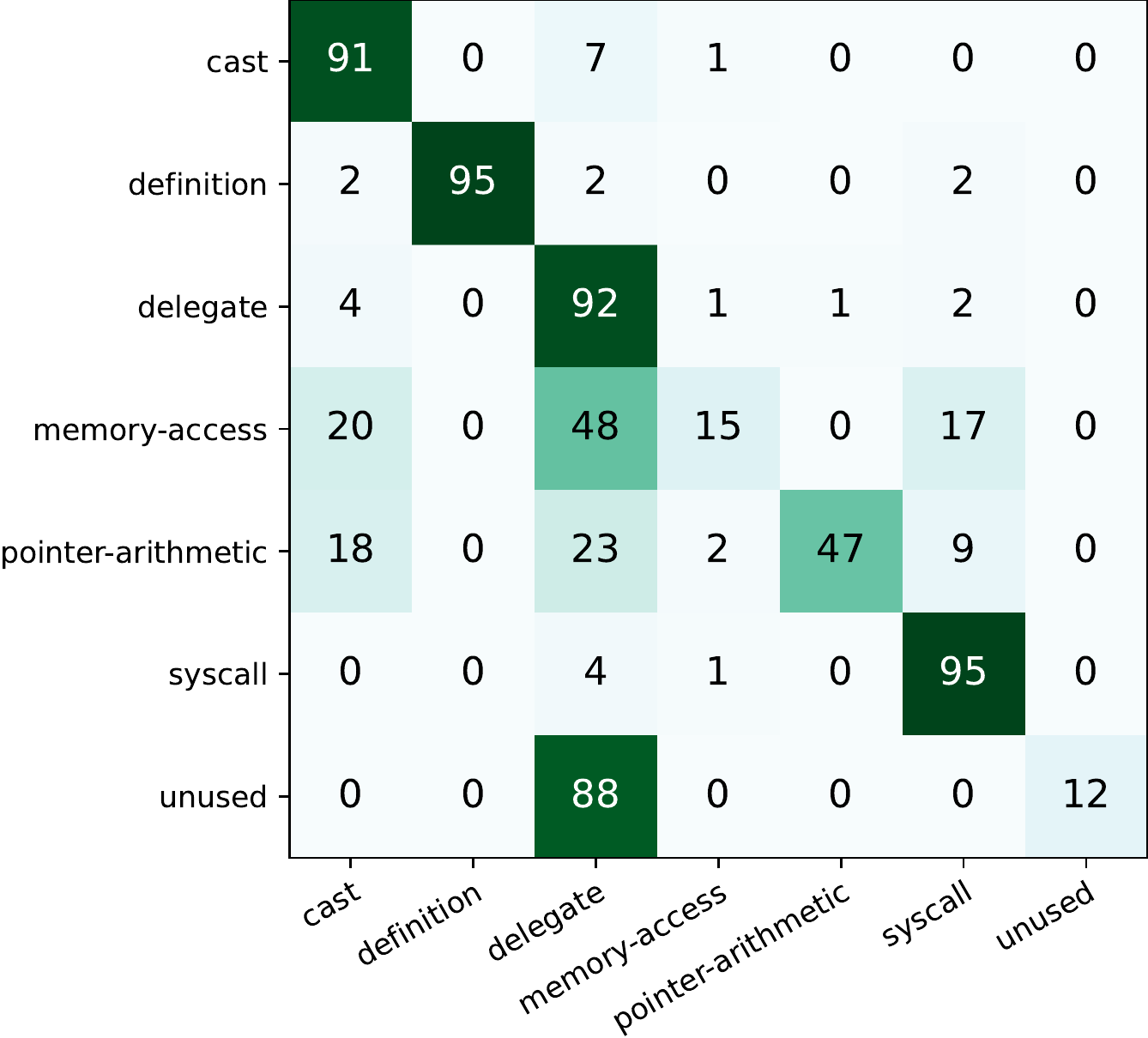} &%
        \confusionMatrix{\labelIwidth}{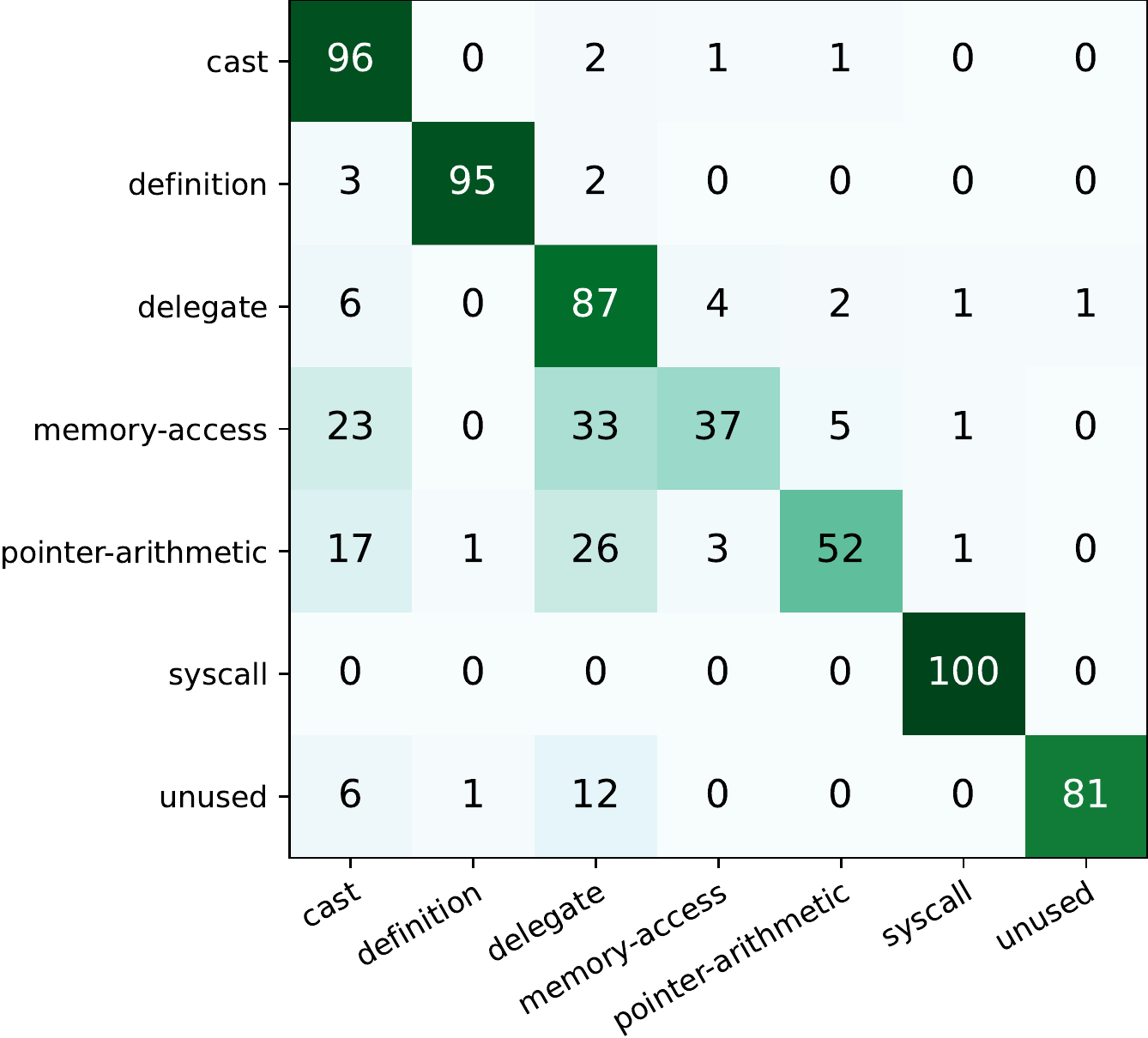} &%
        \confusionMatrix{\labelIIwidth}{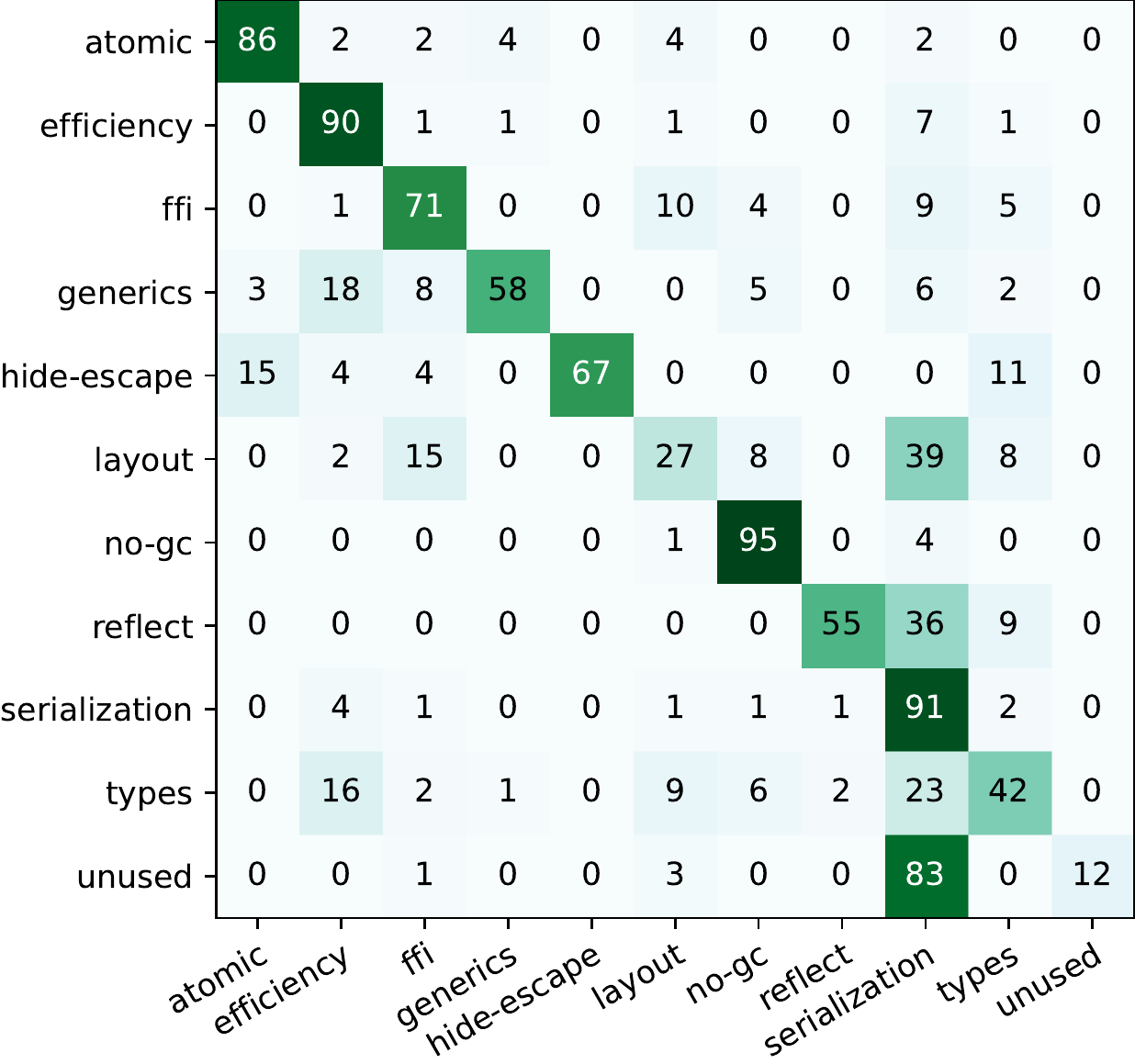} & %
        \confusionMatrix{\labelIIwidth}{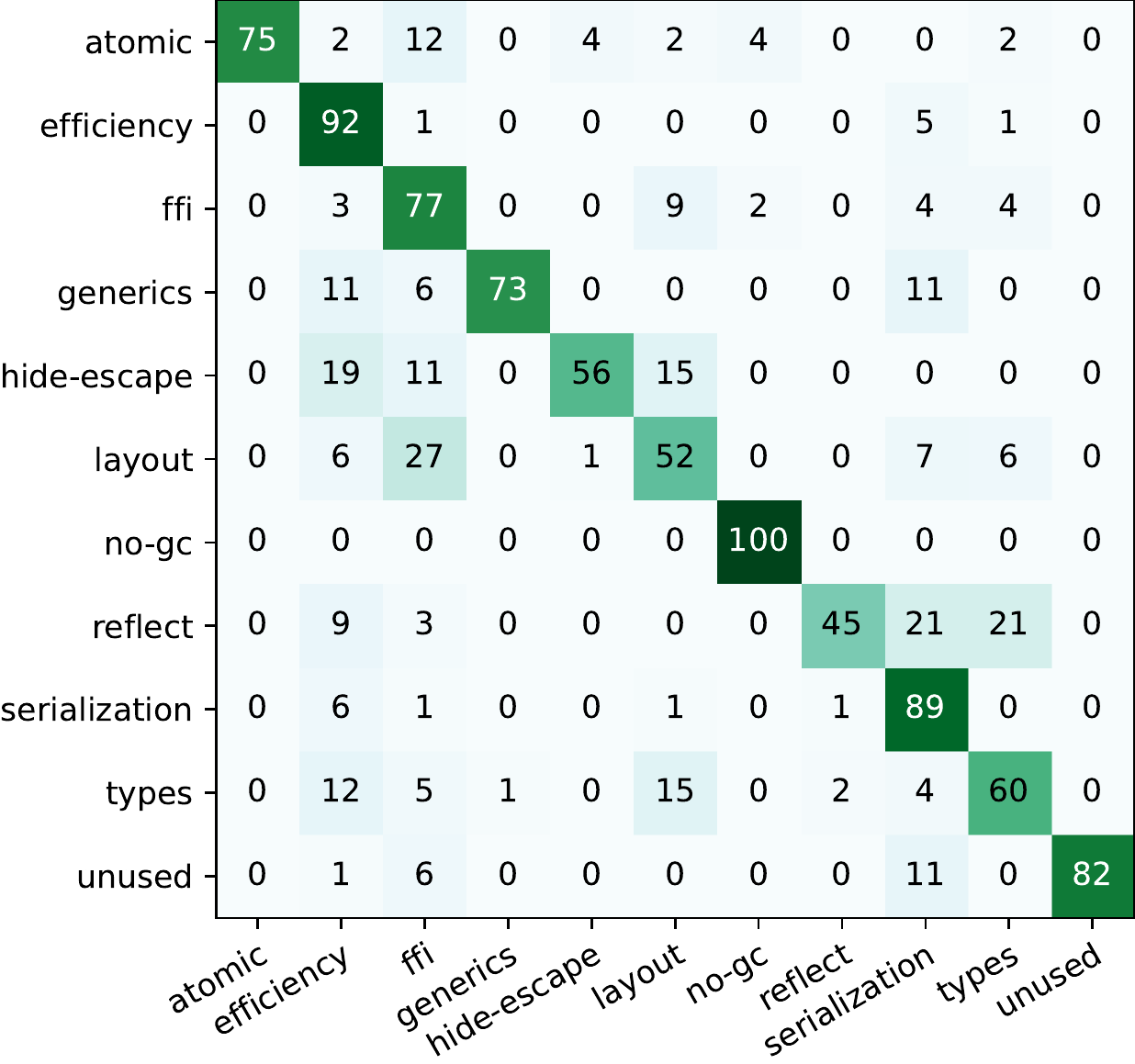}
	\end{tabular}
    \caption{
        Row-normalized sum over the confusion matrices on the test splits using the complete set of vertex features.
        The horizontal axis corresponds to the predicted labels, the vertical axis to the true labels.
        Values are given as percentages.
    }\label{fig:eval:conf-mat}
\end{figure*}

\subsection{Example Prediction: \cref{lst:unsafeclassificationexample}}
\label{sec:eval:examplepred}

We elaborate on the predicted labels for the example presented in \cref{lst:unsafeclassificationexample} to gain insights into concrete predictions. 
We consider the labels (\lOne{}: \textit{pointer-arithmetic}, \lTwo{}: \textit{serialization}) defined by Lauinger et al.~\cite{lauinger2020uncovering} as the expected ones and use the term alternative labels to refer to the two additional labels (\lOne{}: \textit{cast}, \lTwo{}: \textit{efficiency}) discussed in \cref{sec:exampleusage}. 

The baseline \ac{mlp} model, which only considers the single statement vertex containing the \unsafe{} usage that is to be classified, successfully predicts the expected labels for \lOne{} and \lTwo{} as its top-1 predictions. 
Within the conformal prediction, the alternative label for \lOne{} has the second-highest probability. 
However, the alternative label for \lTwo{} is not included in the conformal set. 
We attribute this to the fact that the information that hints at \textit{efficiency} is not present within the single vertex feature vector given to the \ac{mlp} classifier.

The top-1 prediction of the DeepSets and \ac{gin} models include an expected label and an alternative label for \lOne{} and \lTwo{}. 
Both of these labels for \lOne{} and \lTwo{} are included in the top-2 predictions for the models. 
For DeepSets, we observe a larger set size than those predicted by \ac{mlp}. 
We attribute this observation to the additional information that slightly misled the DeepSets model.
For example, for \lOne{} the data type \mintinline{go}{bool} used in Line~\ref{lst:l:typebool} is a positive indicator\footnote{It is among the top-3 important features returned by $\textrm{Grad}\odot\textrm{Input}$~\cite{SanchezLengeling2020}.} for two of the predicted labels, namely \textit{cast-basic} and \textit{cast-bytes}.
As the model is unaware of the control flow, it cannot distinguish if the type influences the \unsafe~usage.

The \ac{wl2gnn} model returns both labels for \lOne{} and \lTwo{} as its top-1 prediction. 
Additionally, this model was the only one predicting no further labels in its conformal sets.

\subsection{Applications of \ourtool{}}
\label{sec:eval:applications}

We discuss two exemplary applications of \ourtool{}: (a) helping security auditors to identify fragments with security-relevant \unsafe{} usages and
(b) assisting developers in refactoring code to replace usages of \unsafe{} that mimic generics with the generics language construct recently added to Go.

We use the confusion matrices in \cref{fig:eval:conf-mat} to estimate the effectiveness of \ourtool{} when applied to the above use cases. 
These matrices present the recall of each label (in the diagonal) and the percentage of falsely predicted labels (other values in the row of the label). 
We present the confusion matrices for the \ac{mlp} and \ac{wl2gnn} models;
the \ac{wl2gnn} model is considered as a representative of all (similarly performing) context-aware models and the \ac{mlp} is the non-context-aware model.

\subsubsection{Security Audit} 

When conducting a security audit, 
\ourtool{} can help auditors to prioritize code fragments that require a manual review by focusing their attention on the most security-relevant \unsafe{} usages, i.e., usages labeled as \textit{casts}, \textit{pointer-arithmetic}, and \textit{memory-access} in \lOne{}. 

The top-1 recall for the \ac{mlp} model to classify \textit{memory-access} and \textit{pointer-arithmetic} is 15\% and 47\%; the \ac{wl2gnn} model achieves a much better recall of 37\% and 52\% -- hence, we focus on this model in the following. 
As already stated, \ac{wl2gnn} predicts \textit{memory-access} with a 37\% accuracy; 5\% of the falsely predicted labels for \textit{memory-access} are classified as \textit{pointer-arithmetic}.
The most confusing label is \textit{delegate} with 33\%. 
For \textit{pointer-arithmetic} we have a 3\% confusion with \textit{memory-access} and 26\% confusion with \textit{delegate}.
Thus, \ac{wl2gnn} narrows down an \unsafe{} finding to \textit{memory-access}, \textit{pointer-arithmetic}, or \textit{delegate} with a combined top-1 recall of 75\% to 81\%. 
As all other labels have quite high detection rates (81\% to 100\%), one can easily rule out other \unsafe{} usages. 
The only other relevant source of confusion are \textit{cast}s, which account for 23\% to 17\%; this is, however, not a problem, as casts are also relevant for security audits. For actual cast operations, \ac{wl2gnn} achieves a recall 96\%.\footnote{It can even distinguish types of casts, e.g., structs, bytes, basics, etc.}

\subsubsection{Refactoring Generics}

Developers can use \ourtool{} to identify \unsafe{} usages that mimic generics (labeled \textit{generics} in \lTwo{}). 
For each usage labeled as such, they can decide if and how to replace it with the newly introduced language feature for generics. 
The recall for label \textit{generics} varies between 59\% for \ac{mlp} and 73\% for \ac{wl2gnn} model. 
While a 73\% recall is already quite high, 
the effective recall for real-world generics detection is even higher. 
In practice, the absence of language-integrated generics is often compensated with code duplication. 
Thus, it is sufficient to detect one of the duplicates to refactor all \unsafe{} usages that mimic generics. 

\begin{summarybox}{Summary}
We conclude that our approach is practical for the discussed 
applications.
Further, \ac{wl2gnn}, \ourtool's default model, performs well in both use cases.
\end{summarybox}

\subsubsection{Generalizability}
In addition to the two above-mentioned concrete use cases, \ourtool{} can be used in other contexts.
The used \ac{gnn} models can cover other types of \unsafe{} usages because our implementation can be used to train them with other Go datasets that contain other types of \unsafe{} usages or other kinds of vulnerabilities.
Further, one can replace or extend parts of our pipeline, such as the implementation to create our code representation, to integrate data sets in other languages.

\section{Threats to Validity}
\label{sec:threats}

A potential threat to \textbf{internal validity} is the correctness of our implementation, evaluation, and \ac{cfg} representation. 
We only included a small subset of the information available for a given \unsafe{} usage and exclude call sites and callees in the \ac{cfg}. 
Thus, not all relationships between different elements are fully represented.
We decided against exploring more fine-grained representations as our experiments to encode the abstract syntax tree of statements did not improve the classification accuracy. 
Further, more information increases the graph sizes and the time required to train a model.
Further, currently the representation is intraprocedural.
Also, we ignore natural language information from comments, even though a human annotator might consider them when labeling a usage. 

The selection of our dataset is a potential threat to \textbf{construct validity}. 
We used the dataset from our previous work~\cite{lauinger2020uncovering} to train and evaluate our classifier.
The dataset by Costa et al.~\cite{costa2021breaking} is not selected as it provides labels only on a file level which is not suitable for our research.
Further, for a fair comparison we did not compare \ourtool{} against the existing static analyzers that can only detect \unsafe{} usages without any classification or only a few usage patterns.

The generalizability of our findings is affected by our threats to \textbf{external validity}.
For training and verification, we rely on the correctness and representativeness of the labels we created previously~\cite{lauinger2020uncovering}. 
As these labels subsume the ones defined by Costa et al.
~\cite{costa2021breaking} and both were derived from usages obtained from popular GitHub projects, we believe that we chose a suitable data set as ground truth. 

Further, our classifier is trained on the \unsafe~package of Go 1.16 and below and does not reflect recent changes in the package. 
Concretely, Go 1.17 (released August 2021) introduced two new functions~\cite{golang2022unsafe} (\cref{sec:unsafepkg}).
These changes were not labeled, as the labels were created before the release. 
However, we believe this will not affect our accuracy as the vast majority of usages are for the type \code{unsafe.Pointer}~\cite{lauinger2020uncovering, costa2021breaking} rather than the \unsafe{} functions.

Lastly, our comparison of different graph convolutions is not comprehensive and only considers existing approaches.
Namely, we chose the DeepSets, \ac{gin}, and \ac{wl2gnn} architectures as representatives of structure unaware, 1-\acs{wl}-bounded, and higher-order convolutions, respectively.
Due to the limited influence of \ac{cfg} edges on the overall model performance, no other types of convolutions were included in the evaluation.
However, we cannot exclude the possibility that a specialized graph convolution operator might improve the prediction quality further. 

\section{Related Work}

\textbf{Unsafe API Usages.} Previous studies on \unsafe{} usages mostly focused on detecting the usages and classifying the usages manually~\cite{lauinger2020uncovering, costa2021breaking}. 
Unlike our work, they relied on a time-consuming, challenging, and error-prone manual validation (see ~\cref{tbl:labels}).
The moderate agreement of the Cohen-kappa score~\cite{mchugh2012interrater} of 0.65 reported by Costa et al.~\cite{costa2021breaking} confirms the difficulty in labeling the different usages precisely.  

Previous studies on \unsafe{} usages in Java focused on detecting usages and their patterns~\cite{mastrangelo2015use} and did not provide a classification tool. 
For Rust, previous studies concentrated on understanding \unsafe{} usages empirically~\cite{evans2020rust, qin2020understanding, astrauskas2020how}.
Furthermore, a survey revealed that most Rust developers use \unsafe, e.g., to use foreign-function-interfaces or interact with hardware~\cite{fulton2021benefits}. 
The RustBelt project provides Rust programmers with formal tools for verifying safe encapsulation of \unsafe{}~\cite{jung_rustbelt_2018}.
Static analyses were developed for a subset of definite \unsafe{} patterns causing bugs~\cite{lauinger2020uncovering,qin2020understanding,bae2021rudra}.
However, the analyses only identify a subset of problems and miss to provide support to classify all \unsafe{} usage patterns.

\textbf{Static Analyses to Detect \unsafe{} Usages.}
Existing analyses for \unsafe{} usages in Go, are either simple linters to detect \unsafe{} usages~\cite{lauinger2020uncovering, securego_G103}, cover only a few selected patterns~\cite{lauinger2020uncovering, go_vet}, or can detect violations only during runtime~\cite{go_cmdcompile}. 
Mature static analyses frameworks such as Doop~\cite{bravenboer2009doop}, Soot~\cite{soot2010}, or Opal~\cite{reif2019judge} do not exist for Go. 
Thus, current analyses are limited to taint analyses~\cite{bodden2016information,li2022cryptogo} or simple AST-based analyses \cite{go_vet, securego_G103}. 
While a taint analysis is not suitable for the task at hand, AST-based analyses are very simple. 
We acknowledge that one could leverage the AST-based analyses to implement a few patterns for categories such as \textit{definition} and \textit{delegate}.
Nonetheless, we doubt that this would work for patterns such as \textit{efficiency} or \textit{reflection} completely. However, we would be interested in future work exploring this line of research..

\textbf{Deep Learning for Vulnerabilities.}
As far as we know, we are the first who classify \unsafe{} usages. 
Thus, we discuss deep learning solutions for vulnerability detection as a related problem.
Recent work on the classification of vulnerabilities, e.g., buffer overflows, has shown the effectiveness of \ac{gnn} architectures~\cite{zhou2019devign, cheng2021deepwukong, chakraborty2021deep, wang_combining_2021, sonnekalb_deep_2022}.
The majority of deep learning classifiers are trained on binary decisions, namely vulnerable or non-vulnerable code ~\cite{sonnekalb_deep_2022,duan_vulsniper_2019}.
Recently, Wang et al.~\cite{wang_combining_2021} predicted the vulnerability type and showed that the precision drops compared to a binary classifier. 
Our classifier achieves a top-1 accuracy similar to their top-3 accuracy. 

Although many approaches for vulnerability prediction support developers and auditors, many provide a coarse-grained prediction level such as file, function, or method-level~\cite{zhou2019devign, sonnekalb_deep_2022, chakraborty2021deep}.
Thus, solutions to reduce the number of lines to be inspected are proposed, e.g., by inspecting the subgraph that influences the prediction the most~\cite{li_vulnerability_2021}. 
Recently, more fine-granular predictions on line-level are suggested~\cite{li_vulnerability_2021}.
However, Li et al.~\cite{li_vulnerability_2021} tokenize code rather than building graphs and leveraging different \acp{gnn} architectures. 

As previous works~\cite{wang_combining_2021, duan_vulsniper_2019} that use \acp{gnn}, we build our intermediate representation by parsing a graph from source code enriched with relevant information for our classification. 
As we aim to solve a different problem, we have to include additional information, such as data types, instead of \textit{LastUse} information~\cite{wang_combining_2021}.
In contrast to Duan et al.~\cite{duan_vulsniper_2019}, our feature vector with 594 possible vertex labels is more expressive than their proposed 144 labels. 

\section{Conclusion}
\label{sec:concl}

In this paper, we presented the first classifier, \ourtool, for \unsafe~usages in Go. 
\ourtool{} helps to understand the actual usage~(\lOne) and the underlying purpose~(\lTwo) of using this escape hatch from memory safety. 
We encoded the \unsafe{} code snippets as enriched \acp{cfg} and classified them with \acp{gnn}. 
To further understand the relevance of features, we varied the included features, e.g., by only including the variables or only the datatype information. %
With the full set of features, we achieve a mean top-1 accuracy of about 88\% and 87\% for \lOne~and \lTwo, respectively, with the \ac{wl2gnn} architecture. 
Furthermore, we show that a set-value conformal prediction classifier returns on average 2 labels with a mean accuracy of 93\%. 
Thus, our classifier is suitable to effectively support developers and auditors to identify and refactor \unsafe~usages, e.g., to replace \unsafe{} with generics or avoid potentially vulnerable \unsafe{} usages. 

In future work, \ourtool{} can be leveraged for automatic large-scale refactoring and auditing tasks.
Further, our methodology and insights gained can be transferred to other domains, such as API misuses in general, by adapting the data set and our implementation.
In addition, our classifier can be used for a comparison of static analyses that can detect the discussed \unsafe{} usage patterns.

\section*{Acknowledgment}
\addcontentsline{toc}{section}{Acknowledgment}

We are grateful for the valuable feedback that we received from the reviewers that helped us to improve the paper. 
We want also to thank Antonio Zhu for his work on wrapping two of our tools into \ourtool{}. 

This paper is based on work funded by the Deutsche Forschungsgemeinschaft (DFG, German Research Foundation) – SFB 1119 – 236615297 and by the German Federal Ministry of Education and Research and the Hessen State Ministry for Higher Education, Research and the Arts within their joint support of the National Research Center for Applied Cybersecurity ATHENE.

\bibliographystyle{IEEEtran}
\bibliography{literature_fixed}

\end{document}